\documentclass[twocolumn,showpacs,preprintnumbers,amsmath,amssymb,floatfix,superscriptaddress,PRC]{revtex4}
\usepackage{graphicx}
\usepackage{upgreek}
\usepackage{tabularx}
\usepackage{dcolumn}
\usepackage{bm}
\usepackage{longtable}

\begin{document}

\title{The $^{63}$Ni(n,$\gamma$) cross section measured with DANCE}  
  
\author{M.~Weigand}
\email{m.weigand@gsi.de}
\affiliation{Goethe University Frankfurt, Max-von-Laue Str. 1, 60438 Frankfurt, Germany}
\author{T.A.~Bredeweg}\affiliation{Los Alamos National Laboratory, Los Alamos, New Mexico, 87545, USA}
\author{A.~Couture}\affiliation{Los Alamos National Laboratory, Los Alamos, New Mexico, 87545, USA}
\author{K.~G\"obel}\affiliation{Goethe University Frankfurt, Max-von-Laue Str. 1, 60438 Frankfurt, Germany}
\author{T.~Heftrich}\affiliation{Goethe University Frankfurt, Max-von-Laue Str. 1, 60438 Frankfurt, Germany}
\author{M.~Jandel}\affiliation{Los Alamos National Laboratory, Los Alamos, New Mexico, 87545, USA}
\author{F.~K\"appeler}\affiliation{Karlsruhe Institute of Technology, Campus Nord, Institut f\"ur Kernphysik, 76021 Karlsruhe, Germany}
\author{C.~Lederer}\affiliation{Goethe University Frankfurt, Max-von-Laue Str. 1, 60438 Frankfurt, Germany}
\author{N.~Kivel}\affiliation{Paul Scherrer Institut, Villigen, Switzerland}
\author{G.~Korschinek}\affiliation{Technical University of Munich, Munich, Germany}
\author{M.~Krti{\v c}ka}\affiliation{Faculty of Mathematics and Physics, Charles University in Prague, V Hole{\v s}ovi{\v c}kách 2, CZ-180 00 Prague 8, Czech Republic}
\author{J.M.~O'Donnell}\affiliation{Los Alamos National Laboratory, Los Alamos, New Mexico, 87545, USA}
\author{J.~Osterm\"oller}\affiliation{Goethe University Frankfurt, Max-von-Laue Str. 1, 60438 Frankfurt, Germany}
\author{R.~Plag}\affiliation{Goethe University Frankfurt, Max-von-Laue Str. 1, 60438 Frankfurt, Germany}
\author{R.~Reifarth}\affiliation{Goethe University Frankfurt, Max-von-Laue Str. 1, 60438 Frankfurt, Germany}
\author{D.~Schumann}\affiliation{Paul Scherrer Institut, Villigen, Switzerland}
\author{J.L.~Ullmann}\affiliation{Los Alamos National Laboratory, Los Alamos, New Mexico, 87545, USA}
\author{A.~Wallner}\affiliation{Department of Nuclear Physics, Australian National University, Canberra, ACT 2601, Australia}
			  
\date{\today}  
\begin{abstract}
The neutron capture cross section of the s-process branch nucleus $^{63}$Ni affects the abundances of other nuclei in its region, especially $^{63}$Cu and $^{64}$Zn. In order to determine the energy dependent neutron capture cross section in the astrophysical energy region, an experiment at the Los Alamos National Laboratory has been performed using the calorimetric 4$\pi$ BaF$_2$ array DANCE. The (n,$\gamma$) cross section of $^{63}$Ni has been determined relative to the well known $^{197}$Au standard with uncertainties below 15\%. Various $^{63}$Ni resonances have been identified based on the Q-value. Furthermore, the s-process sensitivity of the new values was analyzed with the new network calculation tool \textit{NETZ}.
\end{abstract}



\maketitle

\section{Introduction \label{intro}}
\begin{figure}
	\includegraphics[width=0.46\textwidth]{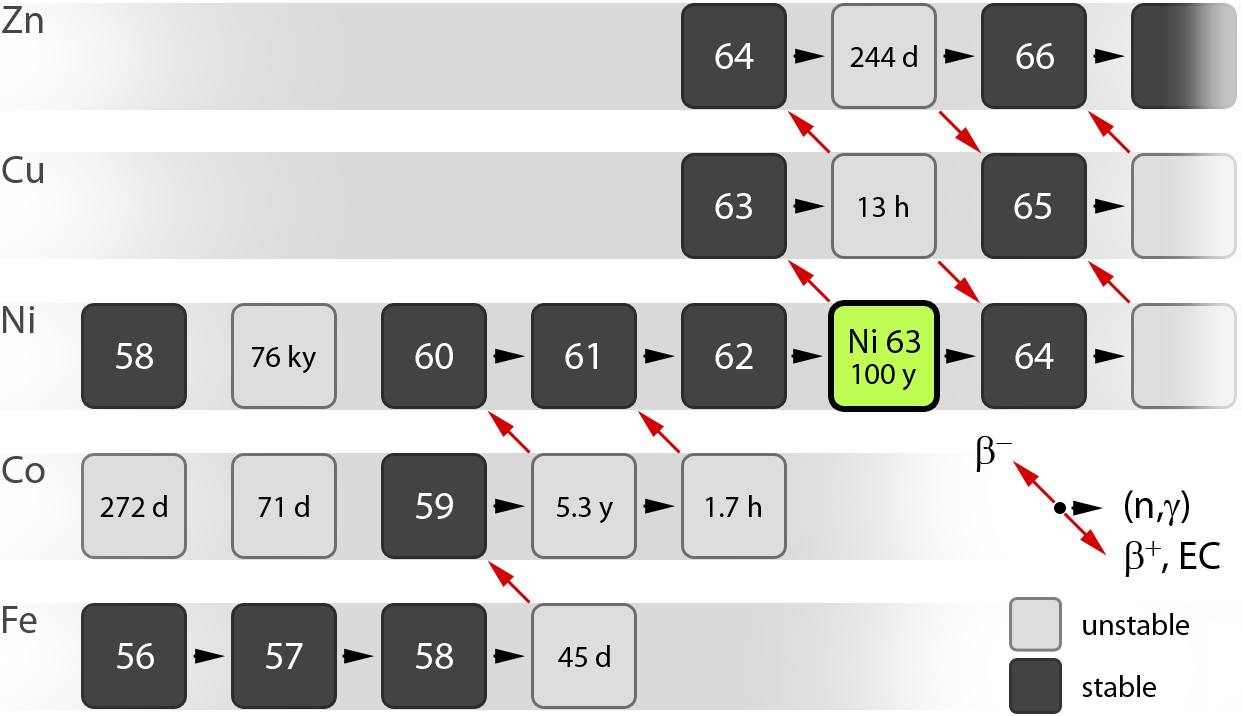}
	\caption{(Color online) The s-process path in the Nickel region, which branches at $^{63}$Ni and proceeds either to $^{64}$Ni, or $^{63}$Cu. The branching ratio depends on stellar temperatures, neutron densities, and the neutron capture CS of $^{63}$Ni. \label{s63}}
	\end{figure}

Nearly all of the observed abundances of elements heavier than iron are either formed by the s- or the r-process in almost equal shares. The specific s-process path depends on temperatures and neutron densities in stars, neutron capture cross sections (CS), and half-lives in case of unstable isotopes \cite{reifarth2014b}. Some of those unstable isotopes in the s-process path play a particular role. They act as branching points when neutron capture and  $\beta$-decay compete, creating different possible ways for the nucleosynthesis mass flow. This branching affects the isotopic abundances of the heavier elements in the s-process \cite{kaeppeler2011,pignatari2010}. Therefore, it is important to know the capture CS for these isotopes. $^{63}$Ni with a half-life of $t_{1/2} \approx$ 100\,yrs is one of these branching points (see figure \ref{s63}). The mass flow can either proceed from $^{63}$Ni to $^{64}$Ni via neutron capture, or via beta decay to $^{63}$Cu.

From the astrophysical point of view there are two scenarios where this branching plays a role: The first are low-mass stars on the asymptotic giant branch (AGB) in a stage of thermal pulses during the He-shell burning phase, where current models consider two neutron sources \cite{herwig2005}. Those are based on the reactions $^{13}$C($\alpha$,n)$^{16}$O and $^{22}$Ne($\alpha$,n)$^{25}$Mg which provide neutron densities of up to $10^{7}$\,$\textrm{n cm}^{-3}$ at a temperature of $10^8$\,K and $10^{9}$\,$\textrm{n cm}^{-3}$ at $3 \times 10^8$\,K, respectively \cite{busso1999,heil2008a}. The second scenario is massive stars in a later stage of development. When they reach the He-core burning phase and later the convective C-shell burning phase, the $^{22}$Ne($\alpha$,n)$^{25}$Mg reaction provides around $10^{6}$\,$\textrm{n cm}^{-3}$ at $3 \times 10^8$\,K and $10^{12}$\,$\textrm{n cm}^{-3}$ at $9 \times 10^8$\,K, respectively.

In the context of the solar abundance pattern, the s-process in massive stars is of major interest, since its signature dominates the nickel region. In the near future the analysis of isotopic nickel abundances in presolar SiC grains will become possible and the $^{63}$Ni CS can also help to constrain the s-process models for AGB stars \cite{stephan2012,stephan2012b}.

Until now, there was only one experimental CS dataset \cite{lederer2013} and theoretical predictions \cite{kadonis2008}. In order to determine the energy dependent neutron capture CS in the astrophysical energy region from $k_BT \approx 0.1$ to 100\,keV, an experiment at the Los Alamos National Laboratory has been performed using the time-of-flight technique and the calorimetric 4$\pi$ BaF$_2$ detector array DANCE, the Detector for Advanced Neutron Capture Experiments \cite{reifarth2004}. Various $^{63}$Ni resonances have been identified via the Q-value. The following sections describe the experimental approach for measuring the $^{63}$Ni neutron capture CS. 


\section{Experiment}
The neutron capture experiment with $^{63}$Ni was performed at the experimental area of flight path 14 (FP14) at the Manuel Lujan Jr. Neutron Scattering Center at the Los Alamos Neutron Science CEnter (LANSCE). The facility offers a white spallation neutron source, driven by a 100\,$\upmu$A pulsed proton beam at a repetition rate of 20\,Hz. 

\subsection{The DANCE array}
	\begin{figure}[b]
	\includegraphics[width=0.46\textwidth]{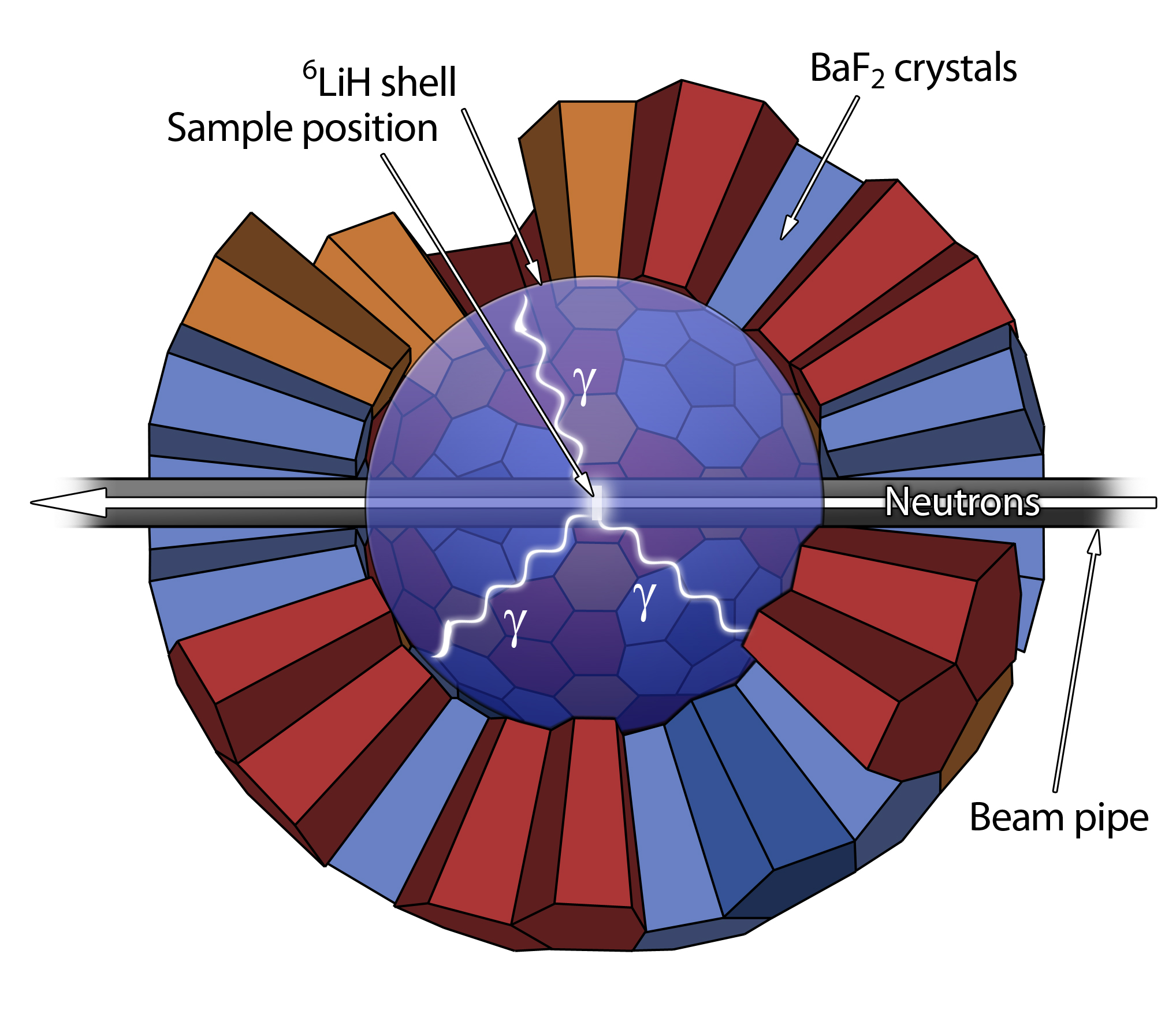}
	\caption{(Color online) A sketch of the DANCE array. Almost half of the crystals are blended out to give an insight. Colors distinguish between different crystal shapes, the blue disk indicates the neutron absorber.}
	\label{baf2}
	\end{figure}
The prompt $\gamma$ rays from $^{63}$Ni(n,$\gamma$) capture events were detected with DANCE \--- a high efficiency detector for $\gamma$ cascades able to run in a calorimetric mode. The array consists of 160 spherically arranged BaF$_2$-crystals (fig.~\ref{baf2}) covering a solid angle of $\Omega$ = 3.6$\pi$ \cite{esch2008}. The full geometry contains 162 crystals, but for practical reasons two have to be left out for the beam pipe. The inner and outer radii of the BaF$_2$ sphere are 17\,cm and 32\,cm. The space between the beam pipe and the crystals is filled with a $^6$LiH spherical shell, which significantly reduces the background induced by scattered neutrons captured in the crystals. This is very important, since capture CS are usually smaller than scattering CS \cite{reifarth2004}.

Summing over all energy deposited from a capture event allows to distinguish between captures on different isotopes based on their reaction Q-values. Additionally, the high segmentation is beneficial when high count rates are expected, and the $\gamma$ ray multiplicity can be used to improve the discrimination of background.

\subsection{Data acquisition and signal processing}
The digitalization of the photomultiplier signals from individual detectors is done with Acqiris DC265 modules with 8\,bit dynamic range. The data acquisition runs with a readout rate of 500\,MHz. Thus, the BaF$_2$ scintillation pulse shapes were sampled in 2\,ns time steps and analyzed right away. Only the time and energy information was stored in order to reduce the data volume, which was done with MIDAS (Maximum Integration Data Acquisition System). A more detailed description of the data acquisition system of DANCE is given in \cite{wouters2006}. Afterwards, the MIDAS data files were processed offline with FARE \cite{wu2012}. With a definable coincidence window, the time information of each gamma detection was used to build capture events within that time window. All events then possess certain properties useful for analysis:
\begin{itemize}
	\item Energy value with long integration time for slow scintillation component of BaF$_2$ ($E_\textrm{s}$), deposited by a $\gamma$ radiated after capture.
	\item Energy value with short integration time for fast scintillation component of BaF$_2$ ($E_\textrm{f}$), deposited by a $\gamma$ radiated after capture.
	\item Total $\gamma$ energy ($E_{\textrm{tot}}$): Sum over all $E_\textrm{f}$ of one event.
	\item Crystal multiplicity ($M_{\textrm{cr}}$): Number of detector modules, that have responded.
	\item Cluster multiplicity ($M_{\textrm{cl}}$): Number of detector module groups, that have responded. Neighboring crystals are combined to clusters.
	\item Neutron energy $E_{\textrm{n}}$ via time-of-flight (TOF) method with a flight path length of 20.29\,m.
\end{itemize}
Further spectral analysis was performed with the ROOT software \cite{brun1997}.

\subsection{Intrinsic Detector Background}
BaF$_2$ scintillators suffer from an intrinsic background because of an unavoidable contamination with $^{226}$Ra and its decay chain via $^{222}$Rn, $^{218}$Po, and $^{214}$Po. They all produce signals, when $\alpha$ particles are emitted and interact with the crystals. In BaF$_2$ crystals $\gamma$ rays deposit about 15\% of their energy in the fast component of the pulse, with a decay time of less than a nanosecond. On the other hand, when an $\alpha$ particle interacts with the crystal, the resulting pulse shape features no fast component. Therefore, by analyzing the scintillation pulse shapes it is possible to identify and discriminate $\alpha$-signals \cite{lanzano1992}.

\begin{figure}[t!]
	\includegraphics[width=0.48\textwidth]{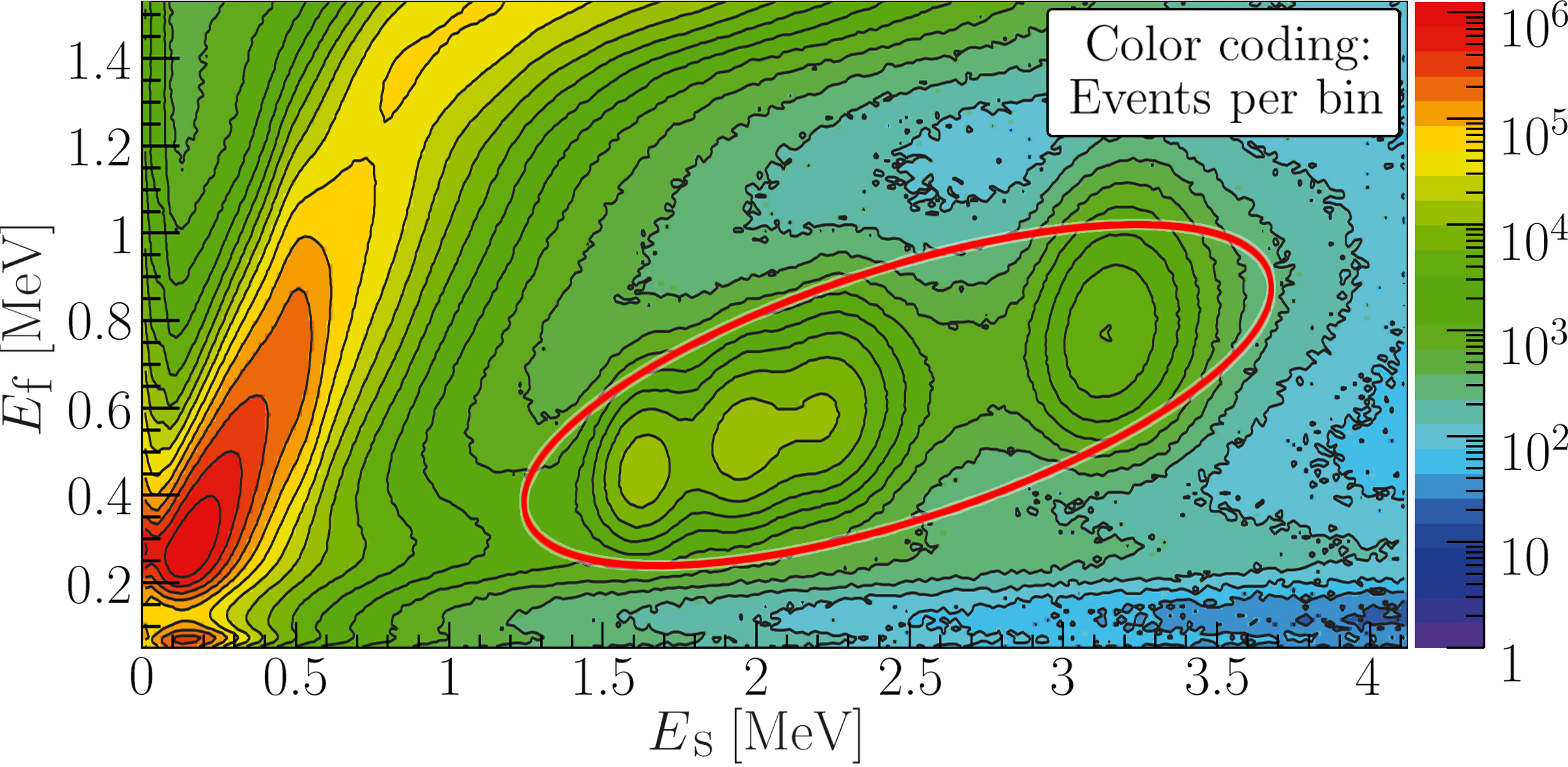}
	\caption{(Color online) 2D plot of $E_\textrm{f}$ versus $E_\textrm{s}$. Signals from alphas seperate in the encircled region.}
	\label{alphas}
	\end{figure}

In the case of the DANCE data analysis, the identification is done via two different integration times leading to the ratio $E_\textrm{f}/E_\textrm{s}$. Figure \ref{alphas} shows the corresponding separation of events from $\gamma$ and $\alpha$ detections, which allowed the almost perfect suppression of $\alpha$-events.

\subsection{The neutron flux at DANCE}
	\begin{figure}[b!]
	\includegraphics[width=0.46\textwidth]{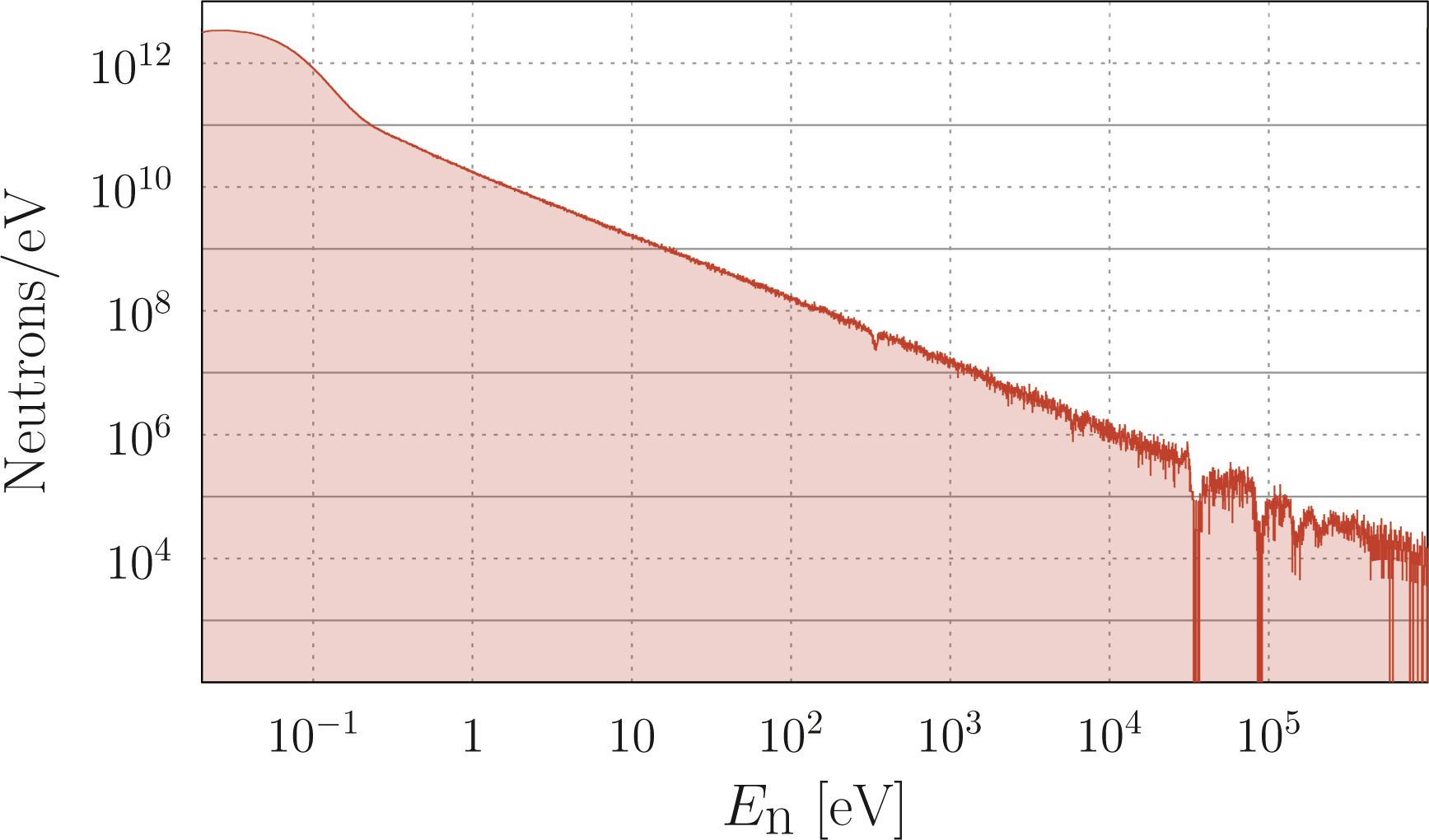}
	\caption{(Color online) The neutron flux at DANCE measured with the $^{6}$Li monitor. The absorption features have their origin in neutron captures in the material of beam line components, Al and Mn in this case.}
	\label{flux}
	\end{figure}
The precise knowledge of the energy dependence of the neutron flux at the position of DANCE is mandatory in order to determine cross sections with DANCE. Therefore, three different neutron monitors measure the flux simultaneously, about 2\,m downstream of the sample position. Their working principle is based on the reactions $^{6}$Li(n,$\alpha$)$^{3}$H, $^{10}$Be(n,$\alpha$)$^{7}$Li and $^{235}$U(n,f). Figure \ref{flux} shows the neutron flux spectrum at the experimental area of DANCE, measured with the $^{6}$Li monitor. Between 1\,eV and 1\,MeV it can be described by a $E^{-1}$-powerlaw. The bump at the low energy end is produced by the water moderator close to the tungsten neutron production target. Several absorption features are visible, which are caused by Mn ($E_\textrm{n} \approx$ 340\,eV) and Al ($E_\textrm{n} >$ 20\,keV), contained in beamline components.

\subsection{The $^{63}$Ni sample}
Since there is no natural resource for the radioactive $^{63}$Ni, it has to be produced. The sample, a metallic foil of enriched $^{62}$Ni with 10.5\,mm in diameter and a total mass of 347\,mg, was exposed to the neutron radiation in the research reactor at the Institute Laue-Langevin (ILL) Grenoble more than 20 years ago \cite{muthig1984,harder1992,trautmannsheimer1992}. This irradiation lasted for 280 days with a thermal neutron flux of $\phi_{\textrm{th}} = 5.5 \times 10^{14}$\,s$^{-1}$ and resulted in a $^{63/62}\textnormal{Ni}$ ratio of $0.108\pm0.002$. The value was determined by means of multiple collection inductively coupled plasma mass spectrometry (MC-ICP-MS). The deployed MC-ICP-MS instrument was a Neptune (Thermo Fisher Scientific, Bremen, Germany), equipped with 9 Faraday cups. The determination was conducted in low resolution mode with a micro-concentric nebulizer at an uptake rate of approximately 50 $\mu$L~min$^{-1}$. 
The sample was subject to chemical separation of the copper produced by the nickel decay prior to the analysis. The total $^{63}\textnormal{Cu}$-amount was estimated as $\simeq6$~mg. Additionally, as the analysis via Q-value and resonance identification showed, $^{59}$Ni ($t_{1/2}$ = 76 kyr) was produced via $^{58}$Ni(n,$\gamma$) during the neutron irradiation as well.

\section{Spectral analysis}
\begin{figure*}
	\centering
		\includegraphics[width=0.98\textwidth]{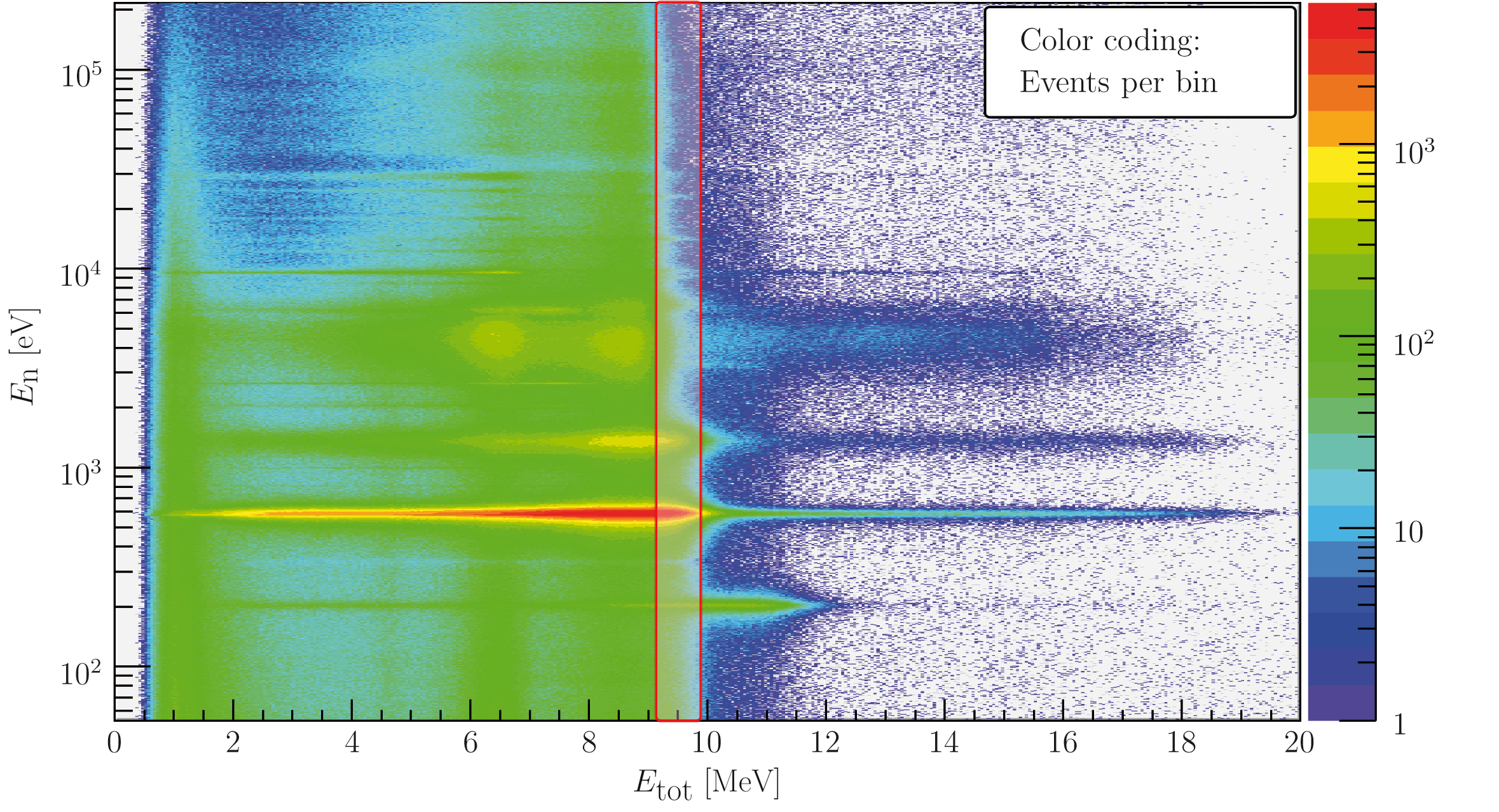}
	\caption{(Color online) A 2-dimensional map of all detected capture events with $M_{\textrm{cl}} > 2$ and with $E_{\textrm{n}}$ on the $y$-axis and $E_{\textrm{tot}}$ on the $x$-axis. Neutron capture resonances appear as horizontal stripes. For the most prominent resonances pile-up effects are visible, but reach only relatively low count numbers. The red box indicates the cut on $E_{\textrm{tot}}$ used for data selection. \label{data}}
	\vspace{1cm}
	\centering
		\includegraphics[width=0.98\textwidth]{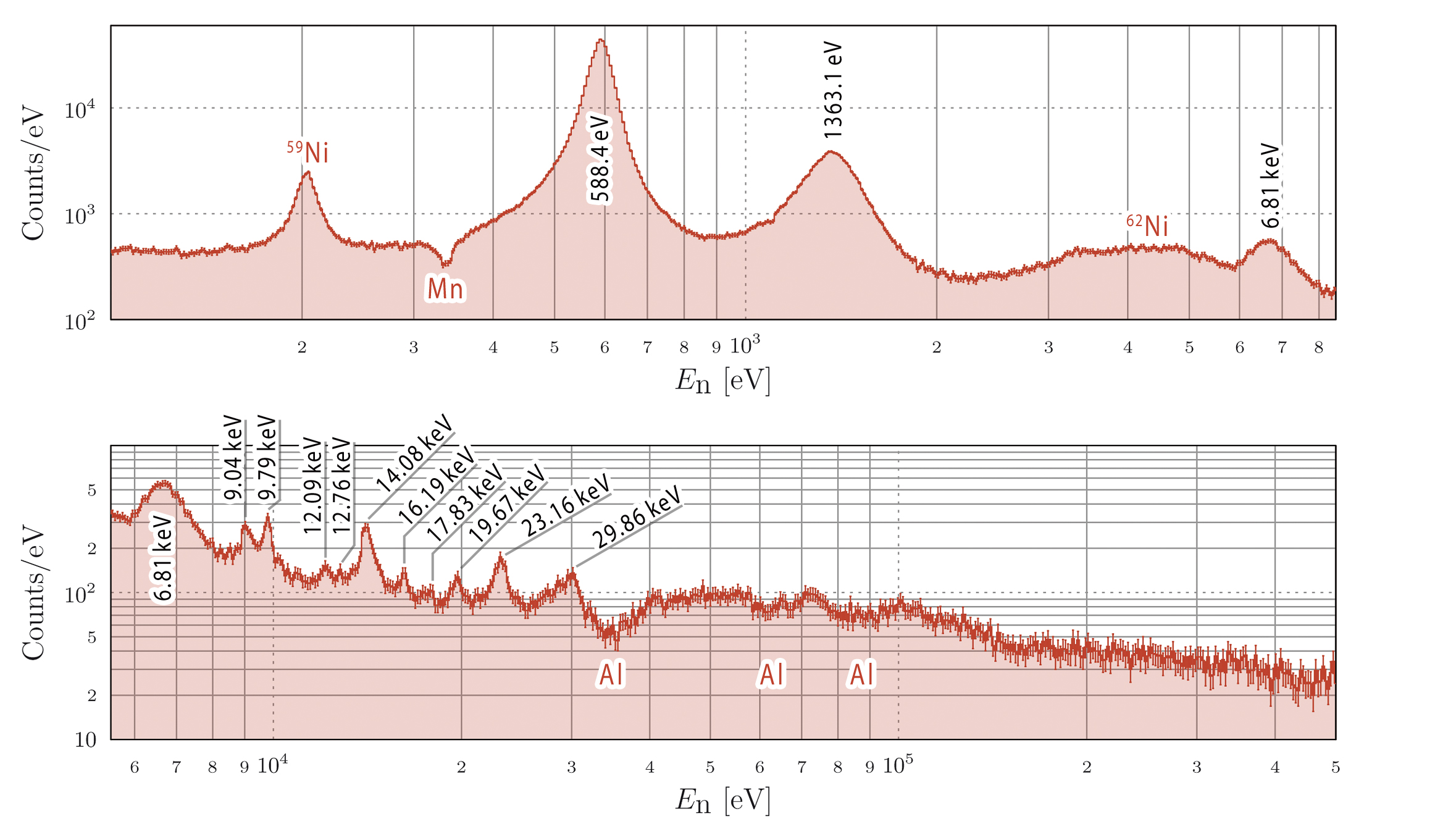}
	\caption{(Color online) The detected capture events per eV between 100\,eV and 500\,keV with an $E_{\textrm{tot}}$ cut from 9.2 to 9.7\,MeV. $^{63}$Ni resonances are labeled with their energies with black font color. Background from other isotopes and absorption features from beamline components (red labels) are denoted as well.\label{details}}
\end{figure*}

Figure \ref{data} shows a 2-dimensional histogram of the number of events as a function of $E_\textrm{n}$ and $E_\textrm{tot}$ obtained with DANCE. The $^{63}$Ni(n,$\gamma$) reaction has a Q-value of $Q_{^{63}\textrm{Ni}} = 9.658$\,MeV. The latter and background from other species in the sample were identified via their Q-values using the total gamma energy information. Also, resonances were assigned to the corresponding isotopes. Two strong and several smaller peaks could be identified as resonances from $^{63}$Ni(n,$\gamma$). They are marked in figure \ref{details}.

Most of the observed background results from neutron captures on the main material of the sample, $^{62}$Ni, and also on the $^{59}$Ni and $^{63}$Cu contaminations. Neutron captures on $^{62}$Ni and $^{63}$Cu have significantly lower Q-values than on $^{63}$Ni: $Q_{^{62}\textrm{Ni}} = 6.838$\,MeV and $Q_{^{63}\textrm{Cu}} = 7.916$\,MeV. $^{59}$Ni(n,$\gamma$),  has a higher Q-value of $Q_{^{59}\textrm{Ni}} = 11.388$\,MeV. But the data show, that a significant contribution can only be found in the area of the strongest resonance at about 200\,eV. 

Additional signals originate from neutrons scattered on the sample which are captured in the detector material, specifically in Ba isotopes. This produces a significant background, which appears at the Q-values of the barium isotopes. $^{137}$Ba(n,$\gamma$) has the highest Q-value with  8.612\,MeV which is close to $^{63}$Ni. However, the contribution of all the contaminants can be reduced with a cut on $E_{\textrm{tot}}$. Figure \ref{data} shows a rough sketch of the applied cut (red box).

\begin{figure}[t]
\includegraphics[width=0.46\textwidth]{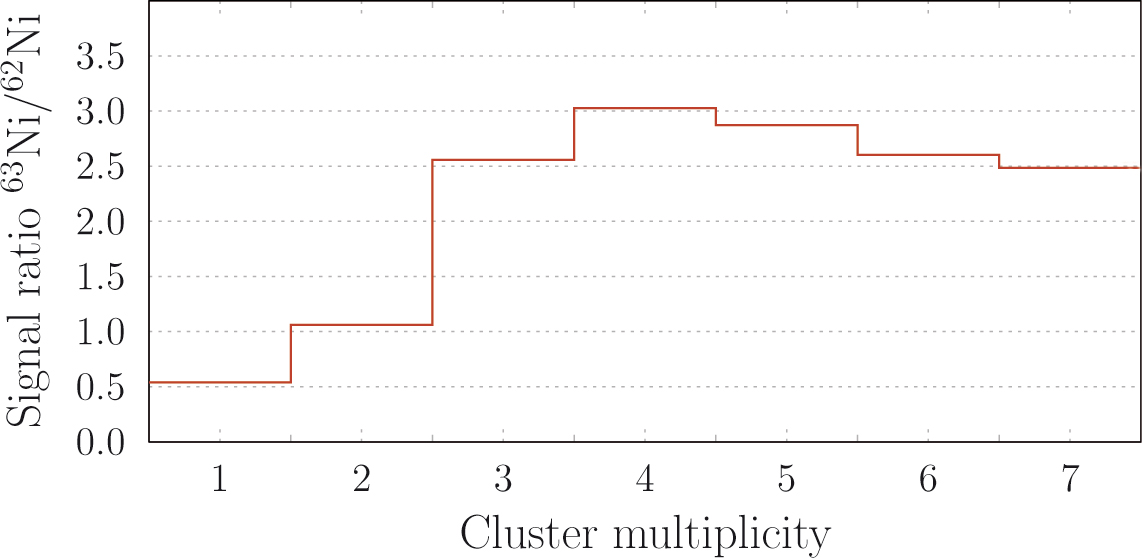}
\caption{(Color online) The ratio of $^{63}$Ni events versus $^{62}$Ni events as function of cluster multiplicity. $^{62}$Ni dominates at low multiplicities, which can be used for data selection and thus, background reduction.}
\label{6362mult}
\end{figure}

Concering $^{62}$Ni, it has also been observed that its capture events exhibit a cluster multiplicity distribution, that is different from $^{63}$Ni. Figure \ref{6362mult} shows the ratio of capture events of both isotopes for each multiplicity. The data were selected by placing cuts on $E_{\textrm{tot}}$ around the corresponding Q-values and on $E_{\textrm{n}}$ around prominent resonances of $^{62}$Ni and $^{63}$Ni. Both distributions were then normalized to 1 to calculate the ratio. The plot shows a higher amount of $^{62}$Ni(n,$\gamma$) events for low multiplicities, which implies an additional tool for background reduction.

\subsection{Background correction}
There are several sources of background during a measurement with DANCE. These are natural radioactivity from the environment and beam-induced background, which can be corrected with measurements with an empty sample holder. Another source is sample induced background with the major contaminant of the sample, $^{62}$Ni, as the dominant source. It is possible to obtain pure $^{62}$Ni dominated spectra based on Q-value cuts and use them for background correction with the following method.

For both Ni isotopes, spectra were obtained by placing a $E_{tot}$ cut around the corresponding Q-values. But, in both cases there are also events that belong to the other isotope. This can be written as two linear equations:
\begin{subequations}
\begin{align}
C_{\textrm{cut}62} &= \epsilon_{62\rightarrow\textrm{cut}63} e_{63} + \epsilon_{62\rightarrow\textrm{cut}62} e_{62} \label{eqn1}\\
C_{\textrm{cut}63} &= \epsilon_{63\rightarrow\textrm{cut}63} e_{63} + \epsilon_{63\rightarrow\textrm{cut}62} e_{62} \label{eqn2}
\end{align}
\end{subequations}

Here, $C_{\textrm{cut}62}$ and $C_{\textrm{cut}63}$ denote the count numbers that result by applying cuts on $E_{\textrm{tot}}$ around the Q-values of $^{62}$Ni and $^{63}$Ni, respectively. In both cases these count numbers are the sum of events from captures on $^{62}$Ni and $^{63}$Ni, represented by $e_{62}$ and $e_{63}$, and weighted by the efficiency factors $\epsilon_{62\rightarrow\textrm{cut}63}$, $\epsilon_{62\rightarrow\textrm{cut}62}$, $\epsilon_{63\rightarrow\textrm{cut}63}$ and $\epsilon_{63\rightarrow\textrm{cut}62}$. Here, e.g. $\epsilon_{62\rightarrow\textrm{cut}63}$ means the efficiency for the detection of $^{62}$Ni(n,$\gamma$) events when a cut around the Q-value of $^{63}$Ni is applied. The equations lead to the following expression for $e_{63}$, which is the number of counts originating only from captures on $^{63}$Ni:
\begin{align}
e_{63} = \frac{{\epsilon_{62\rightarrow\textrm{cut}63}}/{\epsilon_{62\rightarrow\textrm{cut}62}} C_{\textrm{cut}62} - C_{\textrm{cut}63}}{{\epsilon_{62\rightarrow\textrm{cut}63}}/{\epsilon_{62\rightarrow\textrm{cut}62}} \epsilon_{63\rightarrow\textrm{cut}62} - \epsilon_{63\rightarrow\textrm{cut}63}} \mbox{.}
\label{eqn3}
\end{align}

Thus, the efficiency factors had to be determined. First, in order to obtain the efficiency factor ratio $\epsilon_{62\rightarrow\textrm{cut}63}/\epsilon_{62\rightarrow\textrm{cut}62}$, measurements with a $^{62}$Ni sample were performed. After background correction of these data, the ratio was calculated for each $E_\textrm{n}$ bin via the measured counts in the regions enclosed by the $E_{\textrm{tot}}$ cuts. The next sections are dedicated to the determination of the factors $\epsilon_{63\rightarrow\textrm{cut}63}$ and $\epsilon_{63\rightarrow\textrm{cut}62}$.

\subsection{Efficiency determination}
For the determination of the efficiency, not only the gamma cascade detection efficiency, but also the influence of the applied cuts had to be considered. This cannot simply be done with the help of calibration sources. For example, limiting the multiplicity range reduces the efficiency depending on the capturing isotope. However, former works showed that simulations of the detector response of DANCE with GEANT3 deliver results that are on very good agreement with measurements \cite{heil2004}. 
For the deduction of the efficiency, two tasks had to be performed. The first was the simulation of gamma cascades, which describe the decay of highly excited states in $^{64}$Ni. Second, in order to calculate the fraction of the capture events after the application of the cuts, GEANT3 simulations of the complete detector geometry with simulated cascades were conducted \cite{geant1993}. 

\subsubsection{Cascade simulations with DICEBOX}
In order to simulate the response to $^{63}$Ni(n,$\gamma$) events, cascades were generated with the DICEBOX code \cite{becvar1998}. The code uses a Monte Carlo approach to produce levels and decay scheme above $E_{crit}=3.45$ MeV using the statistical model based incorporating various models of level density (LD) and photon strength functions (PSFs).
As we do not know the exact decay pattern of $^{63}$Ni we desided to make simulations with several different combinations of models. In these combinations we used the Constant-Temperature (CT) and the Back-Shifted Fermi Gas (BSFG) models in in the form from \cite{egidy2005,egidy2009} for LD. A parity dependence up to a excitation energy of 5.5\,MeV was assumed in some cases. The Axel-Brink model and model proposed by Kadmenskij, Markushev and Furman \cite{kmf1983} were tested for $E1$ PSF and the single-particle approximation was used for $M1$ PSF. In total 64 dierent model combinations were used. For each combination we generated $10^5$ cascades in each of 15 nuclear realisations \cite{becvar1998}.

\subsubsection{GEANT3 simulation of the detector response}
The DANCE array has been simulated with GEANT3. This includes the following components:

\begin{itemize}
	\item BaF$_2$ crystals
	\item PVC wrapping of the crystals
	\item $^6$LiH neutron absorber shell
	\item aluminum beam pipe
	\item Ni sample and sample holder.
\end{itemize}

\begin{figure}[t]
\includegraphics[width=0.46\textwidth]{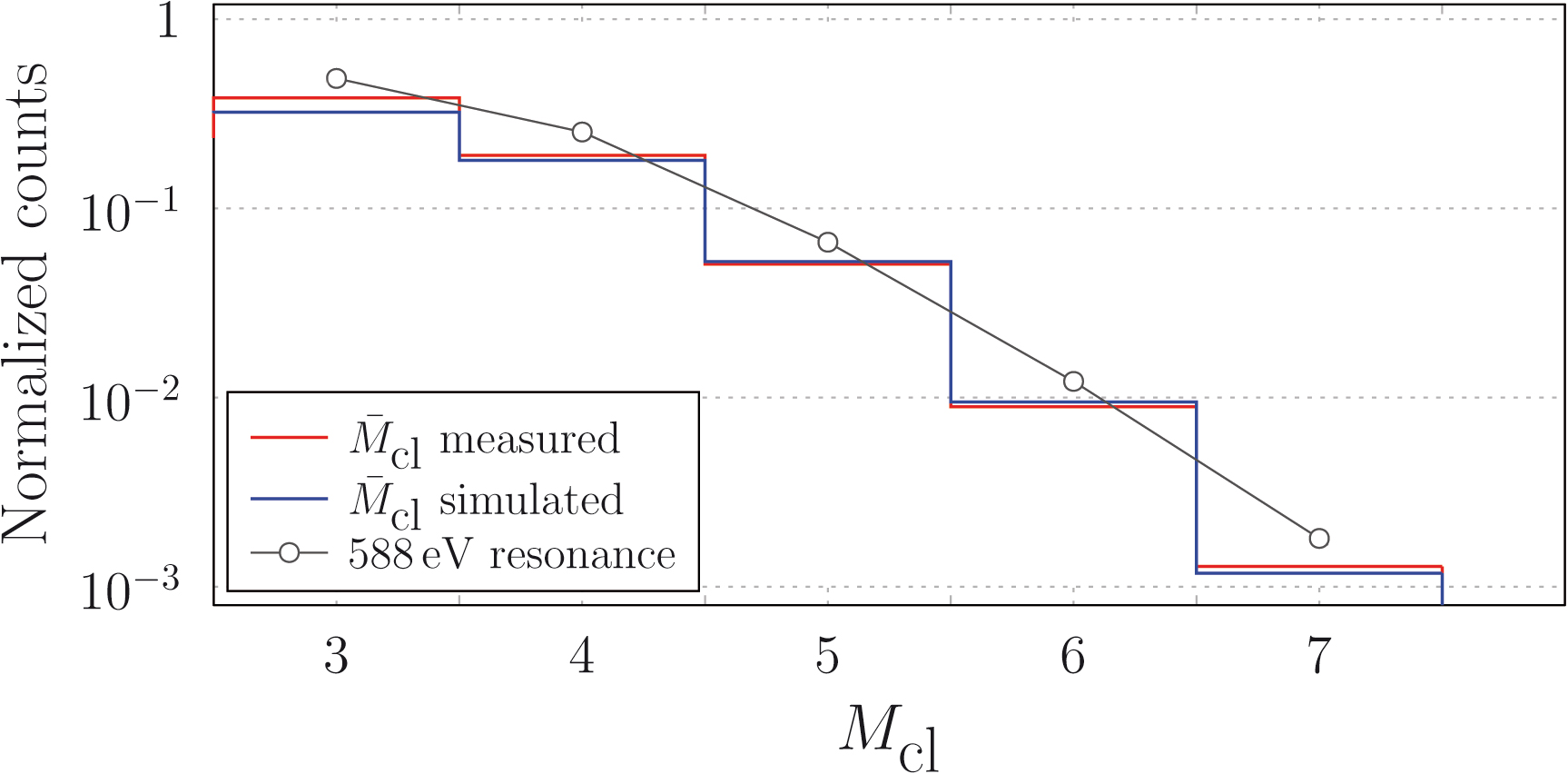}
\caption{(Color online) Comparison of the measured cluster multiplicity distribution (red) with the simulation (blue). The measured distributions of the strongest resonance is depicted with black symbols.
\label{sim_mult}}
\end{figure}
Figure \ref{sim_mult} shows the cluster multiplicity distributions of the simulation in comparisson to the measured mean cluster multiplicity and the strongest resonance. The simulated data are in good agreement with the experimental results. Only the 588\,eV resonance shows a different behaviour. The result of the simulations are the efficiency factors 
\begin{align}
\epsilon_{63\rightarrow\textrm{cut}63} = 0.0653 \pm 4.9\%,\\
\epsilon_{63\rightarrow\textrm{cut}62} = 0.0396 \pm 6.5\%,
\label{eqn3}
\end{align}
for the used energy windows 6.0\,MeV $< E_{\textrm{tot}} <$ 6.5\,MeV and 9.2\,MeV $< E_{\textrm{tot}} <$ 9.7\,MeV, and multiplicities from 3 to 10. The results from all used models were averaged. We give conservative uncertainties for the efficiency factors that contain all the efficiency variations we observe depending on the used settings for the DICEBOX code. Additionally, we doubled the uncertainties for the 588\,eV resonance to take its deviations from the mean multiplicity distribution into account.

\subsection{Scattering correction}
\begin{figure}[t]
\includegraphics[width=0.46\textwidth]{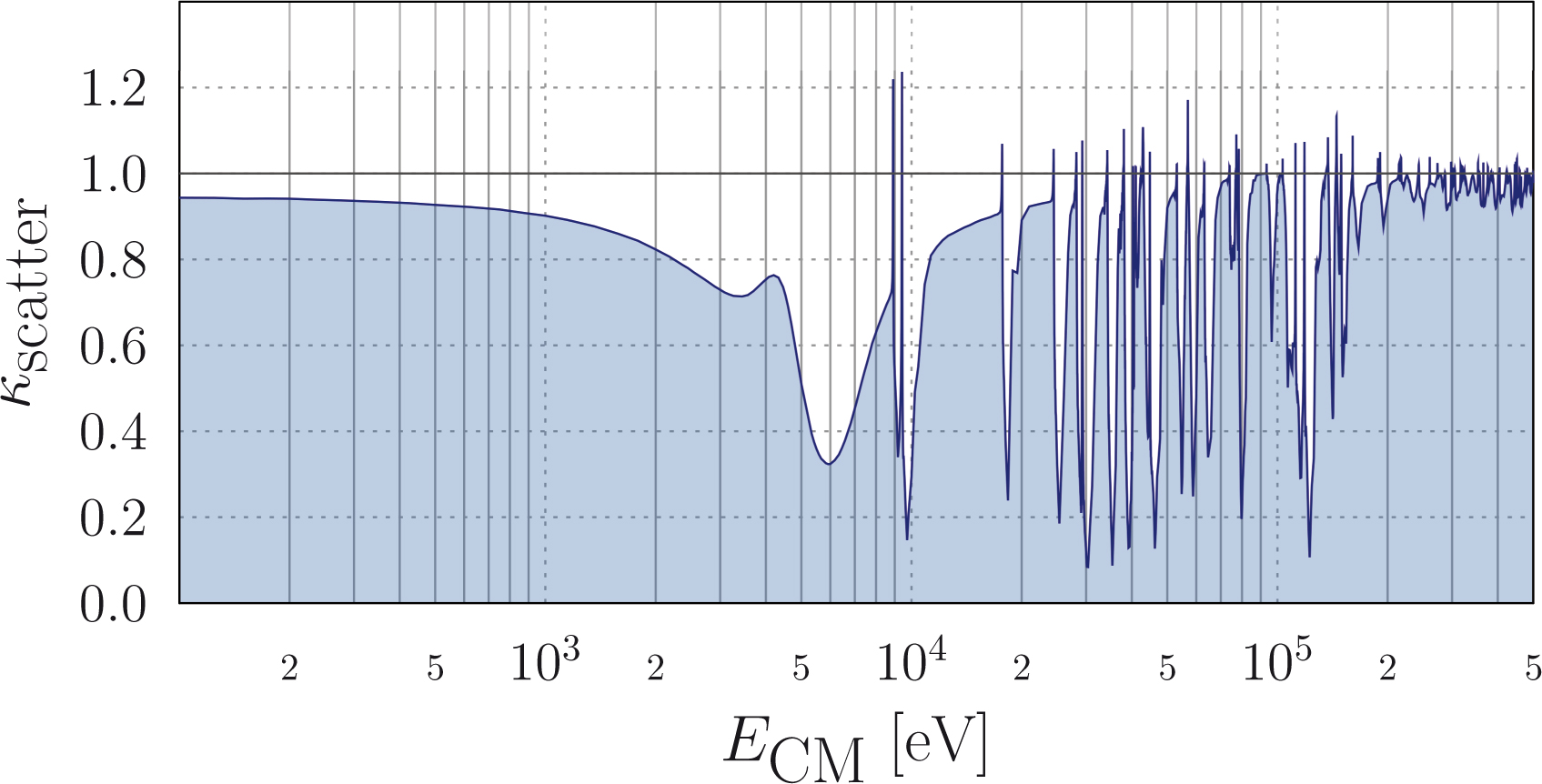}
\caption{(Color online) The energy dependence of the correction factor for scattering in the sample $\kappa_{\textrm{scatter}}$.
\label{sample_corr}}
\end{figure}

For the major contaminant of the sample, $^{62}$Ni, the neutron scattering cross section is larger than the capture cross section. Thus, the effects of neutron scattering in the sample were investigated with a Monte Carlo simulation, based on the CS data available at JEFF 3.0/A. This resulted in an energy dependent correction factor defined as
\begin{equation}
\kappa_{\textrm{scatter}} = \sigma_{\textrm{original}}/\sigma_{\textrm{simulated}},
\end{equation}
and plotted in figure \ref{sample_corr}. When neutrons are scattered in the sample, their mean path length through the sample is extended. This increases the propability for a capture event in the sample and the measured yield is enhanced. The calculated corrections are small for the major part of the measured energy range. But, especially for the broad resonance of $^{62}$Ni between 3\,keV and 10\,keV, the cross section needed significant corrections with a maximal factor of 0.32 at 6\,keV. The statistical uncertainties of the simulation are below 1\%.

This investigation could not be done for the $^{63}$Ni amount due to the fact that there is no measured scattering cross section. But, considering the $^{62}$Ni/$^{63}$Ni ratio of the sample it is expected, that in this case $^{62}$Ni plays the major role. However, new data on the scattering cross section $^{63}$Ni could improve this correction.

\subsection{Normalization}
The well known cross section of $^{197}$Au(n,$\gamma$) was used to normalize the $^{63}$Ni spectra. The CS data was obtained from JENDL 4.0 \cite{shibata2011}. For the measurement, a gold sample with 10\,mm diameter and 500\,nm thickness was used. Due to the small thickness, no corrections for self-absorption neutrons were necessary. The strong resonance at 4.9\,eV was observed with DANCE and used as reference. The relation 
\begin{equation}
\alpha_{\textrm{norm}} = \frac{\sigma_{\textrm{Au}}(E)}{C_{\textrm{Au}}} \frac{\phi_{\textrm{Au}}}{\phi_{\textrm{Ni}}} \frac{N_{\textrm{Au}}}{N_{\textrm{Ni}}} \frac{A_{\textrm{Au}}}{A_{\textrm{Ni}}}
\end{equation}
\begin{figure*}
	\centering
		\includegraphics[width=0.95\textwidth]{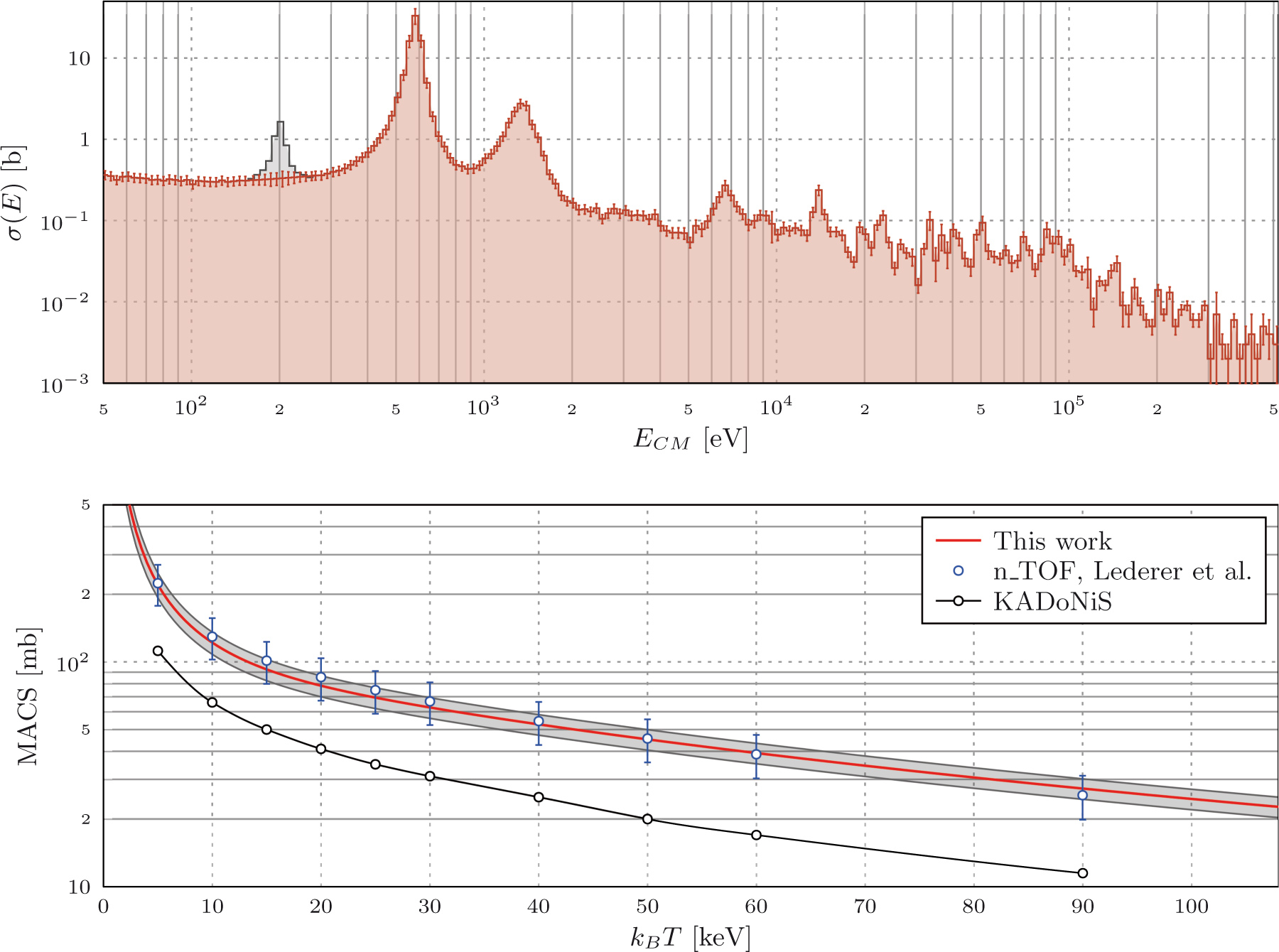}
	\caption{(Color online) \textbf{Top panel:} The neutron capture cross section of $^{63}$Ni as a function of the neutron energy. The grey colored peak at 200~eV was identified as a $^{59}$Ni resonance. \textbf{Lower panel:} The MACS data are shown as red line, the errors are depicted as a grey area. For comparison, data from the KaDoNiS database (black) and from n\_TOF (blue circles) published in \cite{lederer2013} are plotted as well.\label{nickel-macs}}
\end{figure*}
gives the normalization factor $\alpha_{\textrm{norm}}$ for the determination of the $^{63}$Ni CS. The first term is the ratio of the gold CS over the measured number of capture events. $\phi_{\textrm{Au}}$ and $\phi_{\textrm{Ni}}$ are the neutron fluxes for both measurements obtained with the $^{6}$Li neutron monitor, $N_{\textrm{Ni}}$ and $N_{\textrm{Au}}$ are the particle numbers of both samples, and $A_{\textrm{Au}}$ and $A_{\textrm{Ni}}$ denote the cross sectional areas of the samples.

	\begin{table*}[t]
		\caption{Maxwellian averaged cross sections with statistical and systematical uncertainties, and stellar reaction rates in comparisson to the KADoNiS database \cite{kadonis2008} and Lederer et al. \cite{lederer2013}.\label{macsni63}}
		\centering
		\renewcommand{\arraystretch}{1.2}
		\begin{tabular*}{\textwidth}{c @{\extracolsep{\fill}} c c c c c c}
		\hline
		\hline
		 & \multicolumn{3}{c}{KADoNiS} & n\_TOF & \multicolumn{2}{c}{\textbf{this work}} \\
		\cline{2-4}
		\cline{5-5}
		\cline{6-7}
		{\boldmath$k_BT$} & MACS & $\alpha_{\textrm{SEF}}$ & Rate & MACS $\pm$ stat $\pm$ sys  & \textbf{MACS} $\pm$ stat $\pm$ sys & \textbf{Rate} \\
		 $[$keV] & [mb] & - & [cm$^3$/mol/s] & [mb] & [mb] & [cm$^3$/mol/s] \\
		\hline
		\textbf{5} & $112$ & $1.00$ & $6.68\times10^6$ & $224\pm8\pm45$ & $220.62\pm1.78\pm27.65$ & $1.29\times10^7$ \\
		\textbf{10} & $66$ & $1.00$ & $5.56\times10^6$ & $129.5\pm7.1\pm25.9$ & $121.87\pm1.40\pm13.73$ & $1.02\times10^7$ \\
		\textbf{15} & $50$ & $1.00$ & $5.16\times10^6$ & $101.3\pm6.9\pm20.3$ & $92.58\pm1.41\pm10.00$ & $9.44\times10^6$ \\
		\textbf{20} & $41$ & $1.00$ & $4.88\times10^6$ & $85.5\pm6.4\pm17.1$ & $78.33\pm1.38\pm8.28$ & $9.23\times10^6$ \\
		\textbf{25} & $35$ & $1.00$ & $4.65\times10^6$ & $74.9\pm5.9\pm15.0$ & $69.33\pm1.32\pm7.24$ & $9.13\times10^6$ \\
		\textbf{30} & $31$ & $0.99$ & $4.49\times10^6$ & $66.7\pm5.4\pm13.3$ & $62.68\pm1.24\pm6.49$ & $9.04\times10^6$ \\
		\textbf{40} & $25$ & $0.98$ & $4.15\times10^6$ & $54.5\pm4.6\pm10.9$ & $52.71\pm1.10\pm5.42$ & $8.78\times10^6$ \\
		\textbf{50} & $20$ & $0.98$ & $3.69\times10^6$ & $45.6\pm3.9\pm9.1$ & $45.21\pm0.98\pm4.63$ & $8.42\times10^6$ \\
		\textbf{60} & $17$ & $0.98$ & $3.44\times10^6$ & $38.8\pm3.4\pm7.8$ & $39.29\pm0.88\pm4.01$ & $8.01\times10^6$ \\
		\textbf{80} & $13$ & $1.00$ & $3.09\times10^6$ & $29.1\pm2.7\pm5.8$ & $30.59\pm0.71\pm3.11$ & $7.21\times10^6$ \\
		\textbf{100} & $10$ & $1.02$ & $2.73\times10^6$ & $22.5\pm2.1\pm4.5$ & $24.56\pm0.59\pm2.50$ & $6.47\times10^6$ \\
		\hline
		\hline
		\end{tabular*}
	\end{table*}

\section{Discussion and results}
Based on the analysis described in the previous sections, the neutron capture cross section of $^{63}$Ni could be determined from 40\,eV to 500\,keV, which covers the energy range of the s-process. The result is plotted in the top panel of figure \ref{nickel-macs}. Systematic uncertainties arise for different reasons:
\begin{itemize}
	\item The remaining contribution of the $^{59}$Ni	resonance at 200\,eV. Its influence on the MACSs for the s-process however is below 1$\%$ for all considered temperatures. The $^{63}$Ni CS between 150 and 250\,eV was estimated with a linear interpolation as shown in figure \ref{nickel-macs} (top).
	\item The flux normalization is based on the knowledge of the $^6$Li(n,$\alpha$) CS from ENDF. These data were published in \cite{macklin1979}, \cite{lamaze1976}, and \cite{rochman2006}. Its uncertainties are in the order of 6$\%$.
	\item The uncertainties of the efficiency factors are in the order of 3$\%$.
	\item The sample diameter was determined with 5$\%$ and the amount of $^{63}$Ni with 1$\%$ uncertainty.
	\item The Gold CS was used for the normalization and its uncertainties are in the order of 3$\%$.
\end{itemize}
This sums up to a systematic uncertainty of about 7$\%$. Together with the statistic uncertainties, the overall uncertainties stay below 17\%, depending on the number of counts, which is an improvement over the data from n\_TOF.

\subsection{Maxwellian averaged cross section}
Via numerical integration the measured CS was converted into Maxwellian averaged cross sections (MACS) that can readily be used for nucleosynthesis calculations. We provide MACS for thermal energies from $k_BT$~=~5\,keV to $k_BT$~=~100\,keV. The bottom panel of figure \ref{nickel-macs} shows the results compared to the KADoNiS database value which underestimated the CS by a factor of two. Also, a comparison to experimental data from \cite{lederer2013} is given. Our $^{63}$Ni(n,$\gamma$) CS is in very good agreement with the latter. The data are also compiled in table \ref{macsni63}, together with stellar reaction rates $r$ calculated via
\begin{equation}
 r = \sigma_{\textrm{MACS}} N_A \alpha_{\textrm{SEF}} \sqrt{2 E_{\textrm{kin}}/m_\textrm{n}},
\label{eq:rate}
\end{equation}
	\begin{table}[b]
		\caption[$^{63}$Ni resonance data.]{Energies $E_\textrm{r}$ of all $^{63}$Ni resonances that have been identified based on the Q-value.\label{final-res63}}
	  \centering
		\renewcommand{\arraystretch}{1.2}
		\begin{tabular*}{0.46\textwidth}{c @{\extracolsep{\fill}} c c}
		\hline
		\hline
		$E_\textrm{r}$ [eV] & $E_\textrm{r}$ [eV] & $E_\textrm{r}$ [eV] \\
		\hline
		$588.4\pm1.2$ & $12085\pm60$ & $19667\pm47$ \\
		$1363.1\pm3.1$ & $12757\pm42$ & $23159\pm96$ \\
		$6806.4\pm31.7$ & $14078\pm14$ & $29860\pm102$ \\
		$9036.6\pm13.2$ & $16194\pm28$ & \\
		$9787.2\pm17.4$ & $17830\pm54$ & \\
		\hline
		\hline
		\end{tabular*}
	\end{table}

with the Avogadro number $N_A$, the kinetic energy $E_{\textrm{kin}}$, and the neutron mass $m_\textrm{n}$. $\alpha_{\textrm{SEF}}$ denotes the stellar enhancement factor and $\sigma_{\textrm{MACS}}$ the measured MACS. In \cite{rauscher2011} it was shown that the contribution of the ground state to the CS decreases for higher temperatures. But, the contribution of the first excited state is expected to grow in the same way. Thus, $\alpha_{\textrm{SEF}}$ is close to 1 for all given temperatures.

\subsection{$^{63}$Ni(n,$\gamma$) resonances}
The measured differential capture cross section allows to identify $^{63}$Ni(n,$\gamma$) resonances. The determined resonance energies $E_\textrm{r}$ of 13 resonances are compiled in table~\ref{final-res63}.

\begin{figure}[b]
	\centering
		\includegraphics[width=0.48\textwidth]{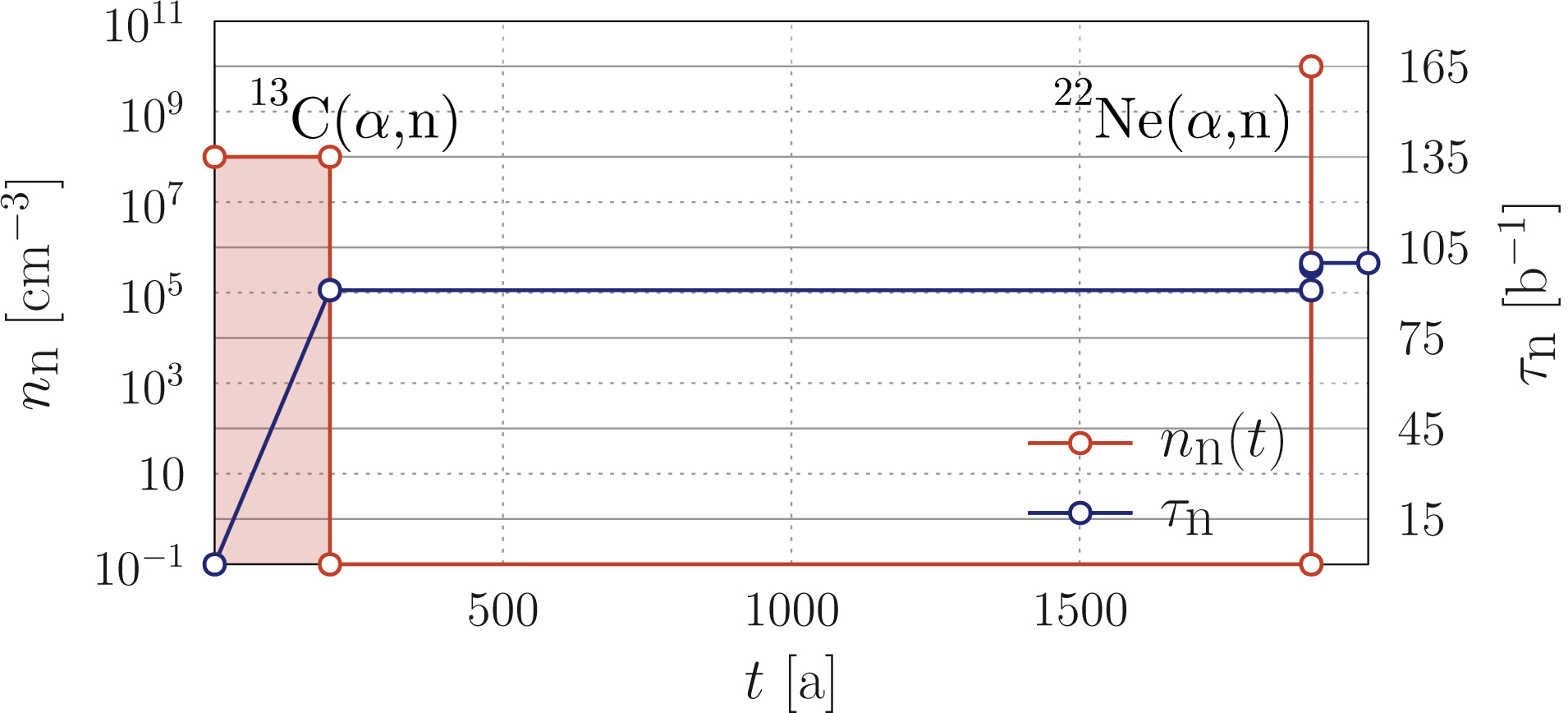}
	\caption{(Color online) The time development of the neutron density $n_\textrm{n}$ and the integrated neutron exposure $\tau_\textrm{n}$ for one thermal pulse in a TP-AGB star. During the inter-pulse phase, the s-process is driven by the $^{13}$C($\alpha$,n) reaction in the $^{13}$C pocket. On the right side the short pulse phase is visible, where the $^{22}$Ne($\alpha$,n) reaction provides the neutrons for the s-process.\label{main}}
\end{figure}

\section{Sensitivity study with \textit{NETZ}}
\begin{figure*}
	\centering
		\includegraphics[width=0.98\textwidth]{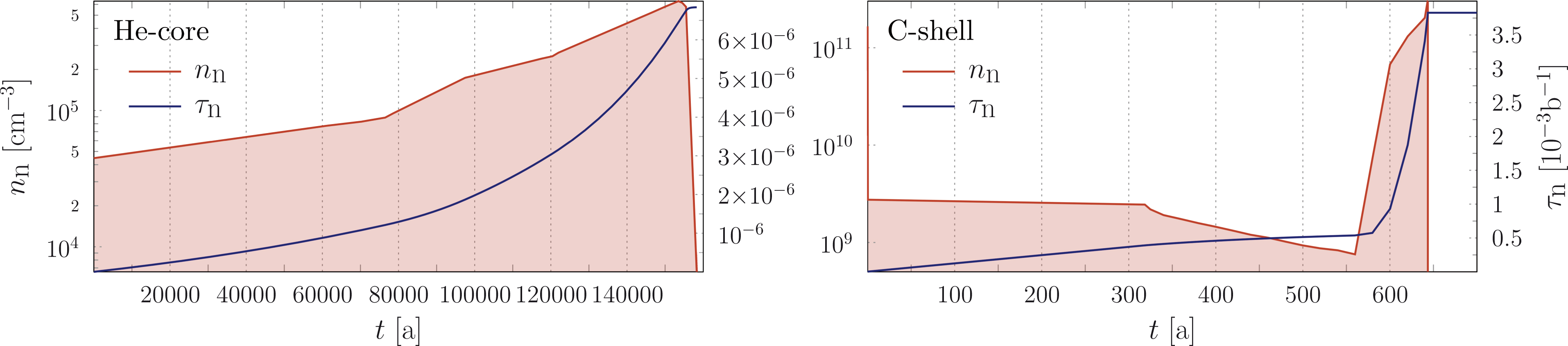}
	\caption{(Color online) \textbf{Left panel:} The time development of the neutron density $n_\textrm{n}$ and the integrated neutron exposure $\tau_\textrm{n}$ for the He-core burning phase in massive stars. \textbf{Right panel:} The C-shell burning phase. In both cases the $^{22}$Ne($\alpha$,n) reaction is the neutron source. The runtime of the s-process during the C-shell burning is much shorter, but the neuton density is higher by orders of magnitude.\label{weak}}
	\vspace{0.8cm}
	\centering
		\includegraphics[width=0.98\textwidth]{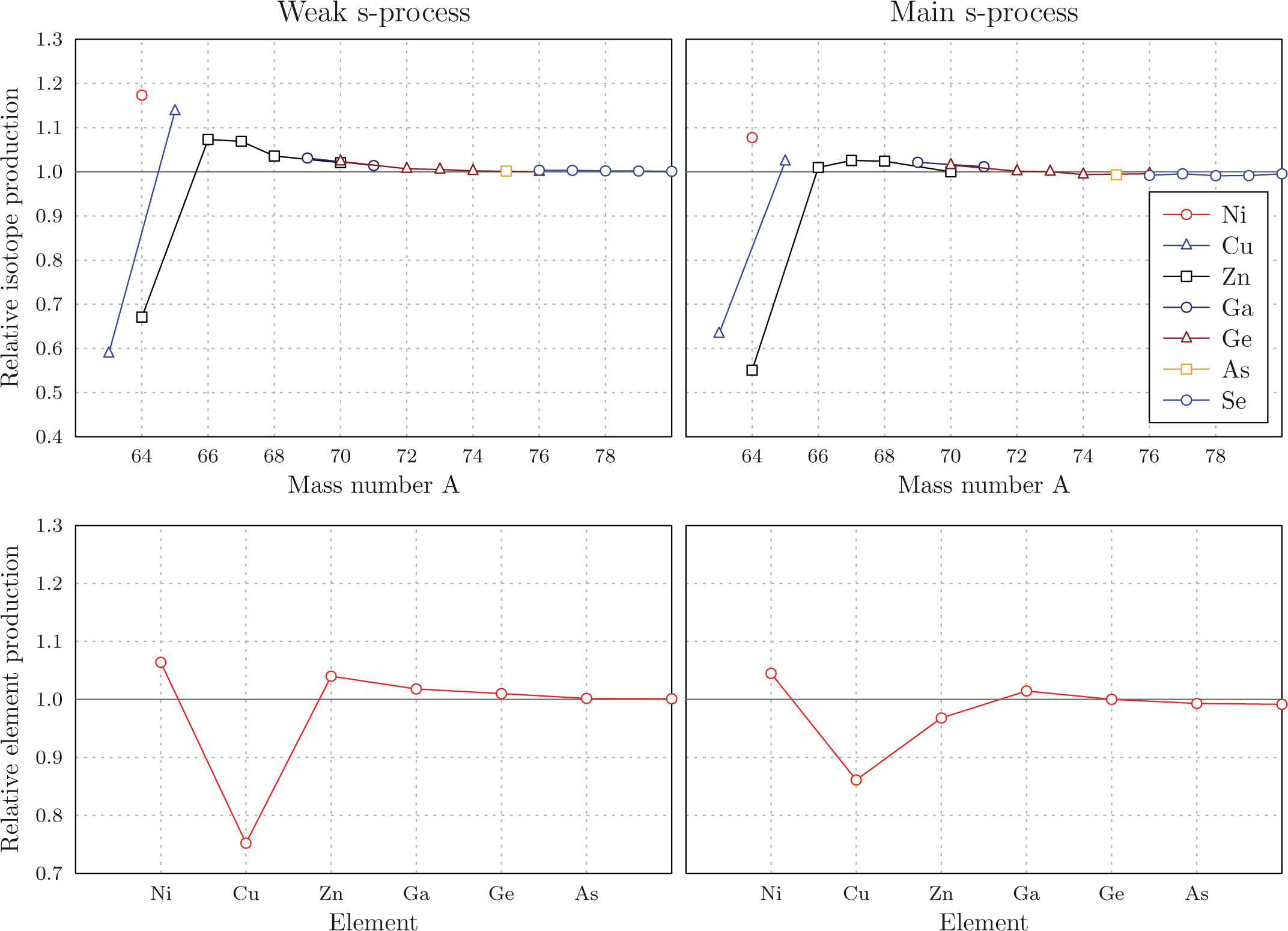}
	\caption{(Color online) \textbf{Upper panels:} The relative production of the isotopes in the vicinity of $^{63}$Ni and lower panels for the weak s-process and the main s-process, respectively. \textbf{Lower panels:} Relative net yields for the elements from Ni to As for both scenarios.\label{netz-results}}
\end{figure*}
Since the neutron capture cross section of $^{63}$Ni is by a factor of 2 higher than expected from previous theoretical predictions, its impact on s-process nucleosynthesis was investigated. This was done with the new and free to use online tool called \textit{NETZ} \cite{jaag2013}. The original code was written by Stefan Jaag \cite{jaag1991} at the Karlsruhe Institute of Technology. Further improvements and the web interface development have been done at the Goethe University in Frankfurt.

\textit{NETZ} is a post-processing tool that uses predefined data for the parameters of the stellar interior and calculates the reaction yields during the given scenarios. Figures \ref{main} and \ref{weak} show the adopted development of the neutron density $n_\textrm{n}$ and the integrated neutron exposure $\tau_\textrm{n}$ for the weak s-process and one thermal pulse of a TP-AGB star \cite{raiteri1993,gallino1998}. Solar abundances from \cite{lodders2005} were used as seed for this study and its relative change indicates the sensitivity on the variation of a particular CS.

As result, the network calculations show a sensitivity on the $^{63}$Ni CS for some nuclei in the vicinity of Nickel. A deviation from the theoretical predictions affects the abundances of $^{64}$Ni, $^{64}$Zn that is not produced by the r-process, and the isotopic ratio of $^{63}$Cu and $^{65}$Cu. Looking at the isotope production relative to the standard case, both s-process scenarios show a similar behavior (see fig. \ref{netz-results}). Since the $^{63}$Ni CS is a factor of 2 higher than previously recommended, the production of $^{63}$Cu and $^{64}$Zn is reduced, whereas the s-process yields of $^{64}$Ni, $^{65}$Cu and the heavier Zn isotopes are enhanced. The net yield of Cu is reduced and exhibits the largest effect, while Ni is slightly enhanced. For Zn the net yield obviously depends on the scenario. In contrast to the weak s-process scenario, the equilibrated mass flow during the s-process in TP-AGB stars allows more mass to pass Zn and reduces its relative production. For elements heavier than Zn, only marginal effects are observed. In \cite{lederer2013} results of a full stellar model simulation for massive stars are presented, that show similar trends.

\section{Summary}
The neutron capture cross section of $^{63}$Ni has been measured in the energy range from 40\,eV to 500\,keV using the 4$\pi$ BaF$_2$ calorimeter DANCE at the Los Alamos National Lab. Additionally, 13 resonances of the $^{63}$Ni(n,$\gamma$) reaction were identified. As a result, we provide new experimental data to improve s-process models. The results are in good agreement with recent data measured at the CERN n\_TOF experiment, published in \cite{lederer2013}. We provide new data with improved uncertainties for network calculations and the refinement of s-process models. The impact of the new CS data on the s-process nucleosynthesis was investigated with the new online tool for network calculations, \textit{NETZ}. The isotopic abundances of Ni, Cu, and Zn are sensitive to the CS change in both scenarios, TP-AGB stars and massive stars. These data will be usefull to constrain the s-process models for AGB stars, when the analysis of isotopic nickel abundances in presolar SiC grains will become possible \cite{stephan2012,stephan2012b}.

\begin{acknowledgments}
This project was supported through the Nuclear Astrophysics Virtual Institute, the Helmholtz Young Investigator project VH-NG-327, and HIC for FAIR. M. Krti{\v c}ka acknowledges support from Czech Science Foundation (grant No. 13-07117S).

\end{acknowledgments}

\bibliography{ref}

\clearpage

\end{document}